# ATHÉNA – A PRE-UNIVERSITY STUDY PROGRAMME AT THE UNIVERSITY OF GENEVA

Andreas Müller*, Madeleine Rousset Grenon**

* Faculty of Science/ Department of Physics and *Institut Universitaire de Formation des Enseignants* (University of Geneva)
andreas.mueller@unige.ch
 **Former Principal of *Collège Claparède* (Geneva) and former member of the *Conférence des directrices et directeurs du Collège de Genève* (2001 - 2017)

**Abstract**
Athena (named after the ancient Greek goddess of wisdom) is a pre-university study programme for mathematics and physics organised by the Faculty of Science at the University of Geneva. It targets pupils enrolled in the final or penultimate year of Secondary II (high school), giving them an opportunity to explore and discover university-level studies in mathematics and physics. The programme aims to enhance pupils' interest for the physical and mathematical sciences by introducing them to new topics, all while giving them a taste for student life. It also seeks to promote scientific careers to young pupils, especially to young women, as well as improving the transition between Secondary II and university.

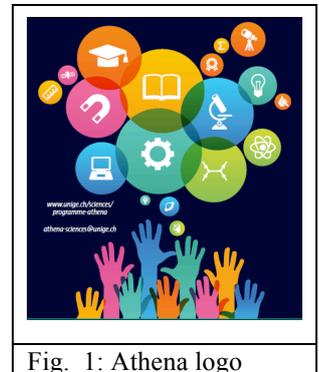

Fig. 1: Athena logo

*This programme really gives a good sense of what studying is like, and prepares us for the student experience in terms of the subject matter itself as well as life on campus.*
(Athena student, 17)

*Even if I don't know what to do after high school, the programme motivated me to look more into physics studies, that door remains open.*
(Athéna student, 19)

## 1    What is Athena?

Young people often display disaffection as well as negative emotions and attitudes towards "hard" sciences at school, and even more so towards careers in STEM fields. These are considerable challenges at both international [1] and Swiss [2] - [4] levels. This has deep repercussions for society. To quote study [4], "there is a lot of work to be done […] to improve the attitudes of pupils, particularly girls, towards scientific subjects, studies and careers". These observations are supported by a report of the Swiss Federal Council, which [2] notes a worrying lack in Switzerland of people trained in the natural sciences, mathematics, and related fields. The report also underlines the particularly low proportion of women in these fields in Switzerland. Fig 2 indicates the number of women relative to the number of men at different stages of education and professional careers (gender ratio, GR). At the end of high school the ratio is considerably greater than 1 (GR = 1.4). Yet it decreases rapidly, for example, in physics studies (GR = 0.32 in Geneva). These data also show that it would be wrong to speak of a general problem in "science", because this disparity is not evident in the life sciences. The Athena programme, organised by the physics and mathematics departments within the Faculty of Science at the University of Geneva, seeks to address this well-known problems by (i) attracting more young people towards studies in these fields, and by (ii) offering targeted support for girls (young women).

Since the academic year 2015/16, high school pupils have been invited to join mathematics and physics classes at the University of Geneva (UNIGE, [5]). This opportunity for early studies is co-ordinated by Camille Bonvin and was initiated by Michele Maggiore and Andreas Müller from the physics department within the Faculty of Science. They drew inspiration from a project in Germany, where a similar programme ("Frühstudium" [5]) has been running for about decades in an increasing number of universities with the support of a large foundation [7].

The main aim is to offer pupils the option to discover and explore mathematics and physics studies in a stress-free environment that is perhaps somewhat off the beaten track and free from self-beliefs an gender stereotypes that prevent young women especially from considering careers in physics or mathematics [2]. The pupils enrolled in the Athena programme choose a course from a pre-determined list to follow during the autumn semester. The pupils are mentored by tutors (undergraduates, postgraduates, doctoral students, or young researchers). In the case of female students, special emphasis is placed on recruiting female tutors. Thanks to the tutors, the programme includes targeted support for the young students, enabling them to develop a more positif and "resilient" self-concept when faced with the intrinsic difficulty of the class content. The tutors also act as role models, known personally to the pupils. The female tutors mentoring female students, in particular, are a living demonstration of how to overcome gender stereotypes.

Each participant who demonstrates serious involvement in the programme receives a certificate of participation. Athena students are also free to pass examinations and, in case of success and subsequent enrolment at UNIGE, can have their course credit recognised.

The Athena programme is in its fourth year a the moment of the writing of this report (for a few statistics, see the bottom of Tab. 1). This is thanks to considerable support from the physics and mathematics departments, the Faculty of Science, and the dean's office at the University of Geneva, as well as to the open and encouraging attitude of the Canton of Geneva's Department of Public Education.

|  | 2015-16 | 2016-17 | 2017-18 |
|---|---|---|---|
| **candidacies** | 110 | 85 | 75 |
| **admissions** | 78 | 74 | 72 |
| **certificates** | 86% | 85% | 79% |
| **examinations** | 45% | 47% | 21% |

Tab. 1. Basic data for the first three editions of Athena (the data for the current academic year 2018/19 are not yet available)

## 2 Bringing an idea to life – practical considerations

Admission: The programme is open to all pupils in public or private schools following a "maturité" (high school) curriculum or any other curriculum that gives access to university studies. Participation is offered completely free of charge to the pupils. Decisions to admit pupils to the programme are based on their grades and teachers' assessment. Consent is also required from the school principal, from their teacher in the subject concerned and, in the case of minors, from their parents or legal guardians. The candidate pupils are informed of the decision in the month of June, for admission in September.

Programme: The pupils selected for the Athena programme follow one or two courses at UNIGE. In principle, the courses are chosen by the pupils (see Tab. 2). The Athena classes take place during school hours. Consequently, the pupils will miss certain hours of their regular high school lessons. However, thanks to active collaboration with school leadership teams, the schedule of Athena classes is designed to have as little impact on high school hours as possible. In particular, the time-tabling constraints are as follows:

**Physics**
– Physics laboratory 1
– Mathematical methods for physicists 1
– Optical Detection of Atmospheric Pollutants
– Electrodynamics 1*

**Mathematics**
– Geometry 1
– Elementary methods
– Probability and statistics*

*advanced; self-test examination needed for entry

Table 2 : Courses available in the Athena programme (example set; there are year-on-year variations)

- The school time-table must be designed in advance, taking into account the inclusion of the university course(s) selected by the pupil.
- Efforts are made to preserve the best interests of the pupil – their strength or weakness in a given subject – when deciding which subjects they will miss; this is determined by the school leadership, in consultation with the time-tabling office.
- Nevertheless, after their pupil's timetables are fixed (often during July), a high school may come to ask the Athena organizers to change the pupil's assigned course that turns out to be incompatible with their time-table. This has almost always proved possible.

School leadership can opt not to release a pupil from a particular class: for example, they typically refuse to release a pupil from laboratory sessions in the elective subjects of biology or chemistry, laboratory hours being very strongly constrained by a lack of available space. This can lead to a pupil not being able to follow the Athena class they had originally foreseen and having to take another instead. Such cases are determined in co-ordination with the Athena organizing team and suitable arrangements have always been agreed upon with relative ease.

In consultation with their tutor, the Athena commission for the relevant department and the leadership of their school, a participant can withdraw from the programme. This decision has no influence on subsequent admission to UNIGE.

Certificates, examinations: At the end of the course, the certificates of participation are distributed at a ceremony dedicated to the Athenians, in the company of their families and teachers. Athena pupils also have the option to sit examinations for university-level course credit. If they pass the exams, the credits thus acquired count towards their bachelor's degrees should they choose to enrol at UNIGE in physics or mathematics. However, in the spirit of exploration and discovery that is at the heart of the Athena programme, participating in examinations is neither expected nor mandatory.

To give the Athena programme a reliable framework, all the procedures have been formalised in a set of organisational rules, and a committee has been set up to carry out the programme.

## 3 Schools' perspectives

The proposal to launch the Athena programme for Genevan high school pupils was immediately and strongly welcomed by the association of principals of Geneva high schools (*Conférence des directeurs/directrices du Collège de Genève*). The idea was perceived as relevant and attractive, and the concept of offering particularly interested high school pupils the opportunity to follow a course or two at university level was deemed to have potential for encouraging the next generation, especially young women. In this way, the programme fits very well in the Canton of Geneva Department of Public Education's action plan for mathematics and natural sciences [8].

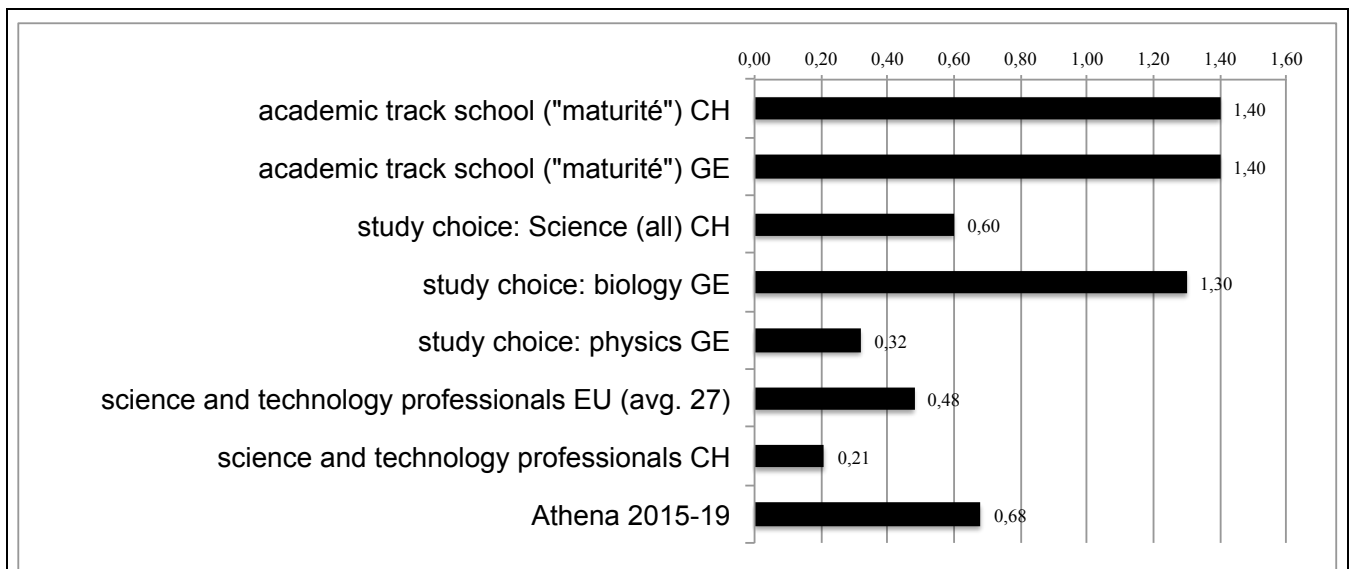

Fig. 2: Ratio of women to men (gender ratio, GR) at different stages of education and professional careers (sources: [2], [9]). Comments:
(i) high school "maturité": young women are a majority! (comparable values at Geneva and Swiss levels);
(ii) study choice, physics: young women "disappear" (from a comfortable majority GR = 1.4 at "maturité" level, to a considerable minority GR = 0.32 for physics);
(iii) study choice, natural sciences/biology: these numbers show that biology, chemistry and physics cannot be grouped as "natural sciences", because GR (as well as many other indicators) is completely different for physics and biology;
(iv) careers: the "disappearance" of women is even more pronounced;
(v) Athena: more than double compared to the value for study choice; young women are interested in an exploratory programme such as Athena in far greater numbers (proportionally) than end up opting to study physics at university (the numbers are similar for mathematics).

For the *Collège de Genève*, it was important to ensure a smooth integration of the programme with the selected pupils' curricula and timetables. The small number of problems that arose were quickly solved thanks to close co-operation between the timetabling offices of the schools involved (see above).

Since the very start of the programme, all pupils wanting to take part and who have received the approval of their school leadership teams have been able to follow Athena courses at UNIGE. Some have stopped part-way through. There have been various grounds for these dropouts: difficulty catching up after a period of illness, course content being very different from expectations, a lack of prerequisite knowledge, *etc*. Nevertheless, a large majority of pupils taking part have successfully completed the programme (see descriptive and evaluative statistics in sections 1 and 4). Indeed, 80 - 90% have obtained a certificate of participation. The rate of withdrawal from the programme ranges around 15 - 20%, which seems – given the exploratory and tentative nature of the programme – completely acceptable. In summary, the Athena programme is very well perceived by high school leadership teams, who seek to support the enrolled pupils as much as possible.

## 4    Evaluation and a few contributions from research

The following evaluation deals with the three editions of Athena in the period 2015/16 – 2017/18 (the data from the current year 2018/19 are not all available yet at the time of writing the present report). The analysis is based on data collected at enrolment, as well as a questionnaire distributed to participants. On average, 70 – 80 pupils per year took part in Athena, with roughly a 50% - 50% split between mathematics and physics, and a gender ratio GR = $N_F/N_M$ = 0.68 (see. Fig. 2). This number is still below that in the target population ("maturité" level overall GR = 1.4), yet it is much higher than for physics studies at university level (for example in Geneva: 0.32). Certain figures (gender ratio, sitting of examinations) display considerable year-on-year variation. This will be a topic of further investigation in future.

The questionnaire, from which a selection of results is presented below, sought to evaluate the Athena programme's objectives in a targeted way. It was adapted from a questionnaire that had previously served to evaluate an early-study programme in Germany [10]. All of the assessment questions used a scale ranging from 1 (least positive opinion) to 6 (most positive opinion). Overall, the pupils have a very positive view of their participation in the Athena programme (see Fig. 3, rightmost column), with an average "grade" between 5.5 and 5.7. There was no gender difference regarding this or any of the following points.

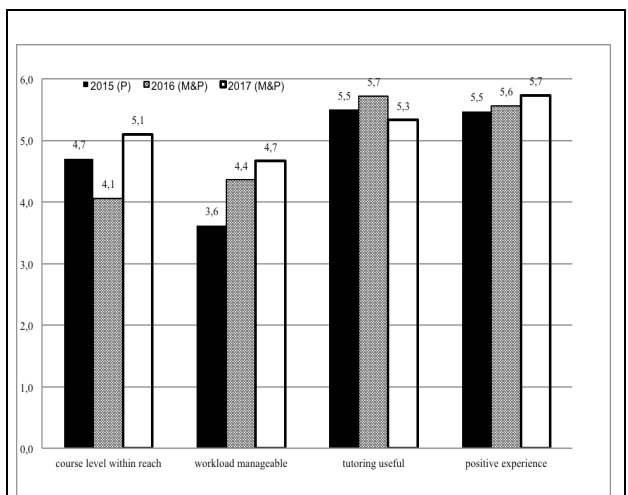

Fig. 3: Perception of the educational offer, by year

Three questions were about the difficulty level and the workload of the courses, and the support available from the tutors as specific characteristics of the educational offer (see Fig. 3). In response to the item "the level of the course was achievable for me", the level of agreement was acceptable (between 4 and 5; it improved by certain adjustments made between the 2016/17 and 2017/18 editions of the programme). The perception of the workload was also improved over the years by organisational adjustments. However, the standard deviations (not presented here) show a considerable variation among the sample: for many participants, the workload is perceived as very heavy. Given these indicators of the workload and the difficulty of the courses, it is encouraging to note the very positive perception of the tutor support.

It is also important to know whether participating in the Athena programme has an impact on the pupils' school results, whether positive (by stimulation) or negative (by overloading). A few questions thus dealt with these aspects (on a scale from -2 = very negative impact, going through 0 = no impact, to +2 = very positive impact). In the subject corresponding to their Athena studies (physics or mathematics), the pupils generally perceive the impact on their school results as positive (+0.8). For other school subjects, the programme is perceived as having almost no impact (-0.1). This result is all the more important given the heavy perceived workload (see above).

These quantitative results are complemented by a few qualitative comments, obtained in an "open answers" section included in the questionnaire, such as those at the start of this contribution, or the examples in Table 3.

| *Atmosphere, perspectives* |
| --- |
| *What pleased me the most is that I now have a better idea of university life: the hours, the atmosphere, the classes, the expectations of the lecturers…* |
| *I appreciated not being under pressure. Learning concepts without any negative repercussions in case of failing the exam made the learning more enjoyable.* (17, f) |
| *The option to discover the system and to understand how the university works – helps avoid making mistakes in choosing a future path. Opens new horizons.* (17, m) |
| *This programme gives a really good insight into what studying is like and prepares us for the university experience in terms of the subject itself and life on campus.* (18, f) |
| *Learning* |
| *The Athena programme is really very interesting. Surrounded by kind students, I learned a huge amount about mathematics and physics.* (16, f) |
| *It is very interesting to discover a maths class at the University of Geneva before being able to come here. This helps me put in place a method for work. In fact, I was able to notice certain mistakes to be avoided in the way I work.* (18, m) |
| Tab. 3: Athena student testimonials |

## 5  Conclusions

In light of the first four editions of Athena, it is clear that is has gone beyond the "proof of concept" stage. Indeed, in 2016 the programme won the Credit Suisse Award for Best Teaching [11]. The Athena programme continues to this day, to the satisfaction of pupils and of the *Conférence des directrices et directeurs du Collège de Genève*. Participants, especially the young women among them, benefit from this opportunity for guidance through its various facets: encouragement (for studies in physics and mathematics, so far), stimulation (the courses and their content, the supplementary talks, the immersion), and welcoming. These will be further enhanced by future modifications to the programme that are currently in preparation.

**Acknowledgements**: We would like to thank Dr U. Halbritter (Cologne) and Prof. E. Niehaus (Landau) for enriching discussions in developing Athena, and Alexander Brown (Geneva) for the translation form French to English.